\documentclass[aps,prd,superscriptaddress,nofootinbib, preprintnumbers]{revtex4}

\usepackage[english]{babel}
\usepackage[latin1]{inputenc}
\usepackage{hyperref}
\usepackage{graphicx,color}
\usepackage{latexsym}
\usepackage{amsmath}
\usepackage{amsfonts,dsfont}
\usepackage{amssymb,slashed}
\usepackage{bbold}
\usepackage{verbatim,epigraph}
\usepackage{float}

\begin{document}

\title{Emergence of Coherent Nanoscale Structures in Carbon Nantotubes}

\author{\textbf{Siddhartha Sen }\footnote{siddhartha.sen@tcd.ie} }

\affiliation{{Centre for Research in Adaptive Nanostructures and Nanodevices}\\ 
\emph{Trinity College Dublin , Ireland}}

\author{\textbf{Kumar S. Gupta}\footnote{kumars.gupta@saha.ac.in} }

\affiliation{{Saha Institute of Nuclear Physics, Theory Division}\\ \emph{1/AF 
Bidhannagar, Kolkata 700064, India}}


\begin{abstract}
Recently unusual properties of water in single-walled carbon nanotubes (CNT) with  diameters ranging from 1.05 nm to 1.52 nm were observed. It was found that water in the CNT  remains in an ice-like phase even when the temperature ranges between 105 - 151 C and 87 - 117 C for CNTs with diameters 1.05 nm and 1.06 nm respectively. Apart from the high freezing points, the solid-liquid phase transition temperature was found to be strongly sensitive to the CNT diameter. In this paper we show that water in such CNT's can admit coherent nanoscale structures provided certain conditions are met. The formation of such coherent structures allows for high values of solid-liquid phase transition temperatures that are in qualitative agreement with the empirical data. The model also predicts that the phase transition temperature scales inversely with the square of the effective radius available for the water flow within the CNT. This is consistent with the observed sensitive dependence of the solid-liquid phase transition temperature on the CNT diameter. 

\end{abstract}
\maketitle

\section{Introduction}

Recently unusual properties of water in single-walled carbon nanotubes (CNT) with  diameters ranging from 1.05 nm to 1.52 nm were observed \cite{mit}. It was found that the water in the CNT  remains in an ice-like phase even when the temperature ranges between 105 - 151 C and 87 - 117 C for CNTs with diameters 1.05 nm and 1.06 nm respectively. Apart from the high freezing points, the solid-liquid phase transition temperature was found to be very sensitive to the CNT diameter, with the freezing point decreasing by $20\%$ for increase of diameter of $.01$ nm. The experimental observations reported are in qualitative agreement with the presence of a single layer of the ice like phase \cite{prev1,prev2}, although the phase transition points obtained  using such a model are significantly different. Neither the sensitive dependence of the transition temperature on the CNT diameter nor its high value can be understood in terms of models considered, as highlighted in \cite{mit}. In view of this alternative ideas need to be considered. In this paper we will explore whether these unusual observed features can be understood in terms of an emergent coherent mesoscopic scale phase of water. The idea of emergent mesoscopic scale coherent structure has been previously used to explain puzzling observed features of mesoscopic magnetism \cite{coey} and of water microbubbles \cite{bubble}. It has been shown that stable coherent mesoscopic structures can form  due to the interaction of bound electrons of the mesoscopic system with the ever present zero-point electromagnetic (ZPEM) field, provided a number of system specific constraints are met \cite{kumar}. Here we suggest that such a picture of mesoscopic coherent surface layer formation \cite{kumar}, which has found applications in mesoscopic magnetism \cite{coey} and water nanobubbles \cite{bubble}, might shed light on the current problem. 

\section{Model}

The physical reason for coherence structure formation \cite{kumar} can be briefly summarized as due to a ZPEM field effecting all bound electrons and lowering their ground state energies in a coherent way \cite{ebbisen}. This idea is demonstrated by considering a mesoscopic system of volume $V$ with  $N$ bound electrons which have  two energy levels and can be described by a Hamiltonian 
\begin{equation}
H = \sum_{i=1}^{N} \hbar \left [ \omega_0 a_0^{\dagger} a_0 + \omega_1 a_1^{\dagger} a_1 + \Omega \left ( a_0^{\dagger}a_1 + a_1^{\dagger} a_0 \right ) 
\right ]_{i} .
\end{equation}
The electron-electron interactions are neglected. For the current problem, the two levels appearing in $H$ are that of electronic states of water molecules which are selected to maximize the possibility of stable mesoscopic coherent structure formation in a CNT. We set the ground state energy to zero. The excitation energy is taken to be $E = \hbar \omega$ and $G \hbar \omega$ is the mixing element due to the ZPEM field between the two bound states where
\begin{equation}
 G=\sqrt{8\pi\alpha}~\sqrt{\frac{c}{V \omega}}~\frac{\delta V}{2V} \rho_{01}.
\end{equation}
Here $\rho_{01}$ represents the mixing of the two energy levels due to the ZPEM field, $V$ is the volume of the mesoscopic surface layer, $\delta V$ denotes the fluctuation of the mesoscopic volume due to the interaction with ZPEM and $\alpha$ is the electromagnetic fine structure constant. The fluctuations due to the ZPEM field are such that $\frac{\delta V}{V} <<1$ (see Appendix A for details). The mixing due to the ZPEM field leads to the lowering of the ground state energy from 0 to $-G^2 \hbar \omega$ and an increase of the excited state energy by the same amount \cite{kumar}.  


A special property of a coherent water layer is that it expels air molecules dissolved in it. The physical reason for this is that molecules of water in a coherent state have a common frequency which leads to an additional time dependent force that is attractive for molecules in resonance but repulsive for those not in resonance \cite{frohlich}. Thus the air molecules not in resonance with the water molecule oscillations, experience a repulsive force, leading to their expulsion (see Appendix B for details).  There are sufficient number of these expelled air molecules to  form a ring surrounding the water molecules \cite{air}. In addition, the hydrophobic surface of the CNT repels the water molecules. In our model we expect that the volume available for water flow is reduced by the presence of the expelled air molecules. Taking the diameter of air molecule to be 2.92 $A^{\circ}$ we expect a gap of this size between the water layer and the CNT surface \cite{reiter,dasilva}. Thus the effective radius $r_e$ of the water column for a CNT with diameter 1.05 nm would be 2.33 $A^{\circ}$ and the corresponding $r_e$ for the 1.06 nm diameter CNT would be 2.38 $A^{\circ}$. It is known that water molecules can form clusters \cite{teeter,fowler,alex,chaplin,johnson} and there is both theoretical \cite{alex,prev1} and experimental \cite{alex} evidence for water clusters in CNT's. Furthermore,  MD simulations of 6-20 water molecules show that most stable clusters are formed by square and pentagonal water structures \cite{mdwater}. Taking the diameter of a water molecule to be 2.75 $A^{\circ}$, the effective CNT volume available for water flow in our model is compatible with the existence of stable water clusters.

The mesoscopic coherent structure of water can only exist and be stable as a quasi 2-dimensional layer \cite{kumar}. In the present case, there are a small number of water molecules in any given cross-section of the CNT, that form the required mesoscopic layer. It was also necessary that the two electronic bound states of the water molecules used in the model are stable.  If these conditions are met,  a coherent mesoscopic layer of water molecules can form as interactions of the surface layer molecules with the ZPEM field results in the decrease in energy of the original ground state energy and  an increase in energy of an excited state that is in resonance with a ZPEM frequency. Such an effect for molecules in a tunable cavity have been observed \cite{ebbisen}. However in \cite{kumar} it was shown that such tuning is not needed for a properly chosen mesoscopic domain volume $V$.  The conditions that the coherent structure be stable at a temperature $T$ are given by \cite{kumar}
\begin{eqnarray}
 kT &<&  \hat{\epsilon} - G^2 \hbar \omega ~\equiv \epsilon \\ 
kT &<& G^2 \hbar \omega
\end{eqnarray}
where $\hat{\epsilon}$ is the difference between the ionization energy of a bound electron in water and the excited state energy of the two level system. 

\section{Results}

We are now ready to state the results of \cite{kumar} for the system of interest. For the coherent layer water volume $V = \pi r_e^2 \lambda$, where $\lambda= \frac{2 \pi c \hbar}{\nu E}$, $E= \hbar \omega$ is the energy difference between the two electronic states of the water and $\nu = 1.5$ is the refractive index of water, we get 
\begin{equation} 
G = 5 \times 10^{-12} \sqrt{\nu}~ \frac{\rho_{01} \sqrt{N}}{r_e^2}.
\end{equation}
The maximum  number of electrons $N$ that can contribute to the coherent structure formation is given by  $N  = \frac{3 c h r_e^2}{4 \nu E r_w^3}$, 

Each water molecule has a volume $\frac{4}{3} \pi r_w^3$ and the coherent domain of water has a volume $V = \pi r_e^2 \lambda^{\prime}$. Thus, the maximum number of water molecules in the coherent volume $V$ is given by
\begin{equation}
 N = \frac{V}{\frac{4}{3} \pi r_w^3},
\end{equation}
where $r_w$ denotes the radius of a water molecule. If the energy difference between the two levels of the system is given by $\hbar \omega = E$, then the corresponding wavelength in the coherent water volume is given by
\begin{equation}
 \lambda^{\prime} = \frac{2 \pi c \hbar}{\nu ~ E} 
\end{equation}
Thus the maximum number of coherent electrons in the volume $V$ is given by
\begin{equation}
 N  = \frac{3 r_e^2 \lambda^{\prime}}{4 r_w^3} = \frac{3 r_e^2 c h}{4 r_w^3 \nu E},
\end{equation}
where $\nu$ is the refractive index of water in the CNT.

The requirement $\frac{\delta V}{V}<<1$ places restrictions on $N$. For the coherent water volume $V$, $\frac{\delta V}{V} = \frac{2\lambda_c}{r_e} \sqrt{N} \sqrt{\frac{2\alpha }{\pi}~\ln{\frac{1}{\alpha^2}}}$, where the Compton wavelength $\lambda_c =  2.42 \times 10^{-10}$ cm and $\alpha = \frac{1}{137}$. The constraint $\frac{\delta V}{V} <<1$ then implies that $\sqrt{N} <<  10^{10} r_e$. Thus $N$ has to be much less than  $5.4 \times 10^4$ and $5.6 \times 10^4$ for the CNT's with diameters 1.05 nm and 1.06 nm respectively. For a typical value of $E= \hbar \omega = 5$ eV, the length of the coherent cylindrical water structure is given by $\lambda = 1.65 \times 10^{-5}$ cm, For a CNT of diameter 1.06 nm, the corresponding maximum number of coherent electrons is given by $N = 2.7 \times 10^{3}$, which falls well within the above limit on $N$.

The parameter $G$ depends on the transition matrix element $\rho_{01}$. Let us estimate its value. $\rho_{01}$ depends on the two electronic states of the water molecule. The two states are assumed to be effective Rydberg states  and carry angular momenta $l$ and $l+1$ respectively, since in this simplified model, the transition matrix between the two states is non-zero only if their angular momenta differ by 1. The geometrical picture uses cylindrical coordinates to locate a water molecule and then describes the effective electronic states in terms of standard spherical polar coordinates. Thus we take 
\begin{eqnarray}
|0> &=&  |l> = C_l r^l ~\exp \left [ -\frac{r}{r_0} \right ] Y_{l,0},~~~~~~~~~~~~~~~~~~C_l = 
\sqrt{\frac{2l+1}{4 \pi \Gamma(2l+3)} \left ( \frac{2}{r_0} \right )^{2l+3}} \\
|1> &=&  |l+1> = C_{l+1}~ r^{l+1}  ~\exp \left [ -\frac{r}{r_0} \right ] Y_{l+1,0},~~C_{l+1} = \sqrt{\frac{2l+3}{4 \pi \Gamma(2l+5)} \left ( \frac{2}{r_0} \right )^{2l+5}}
\end{eqnarray}
which gives
\begin{equation}
 \rho_{01} \equiv \rho(l,l+1) = r_0 \frac{l+1}{2} \sqrt{\frac{2l+4}{2l+1}}
\end{equation}
If we denote by $\epsilon = \hat{\epsilon} - G^2\hbar \omega$, where $E= \hbar \omega$ and  $\hat{\epsilon}$ is the difference between the dissociation scale and the excited state energy of the two level system electronic system, then we get 
\begin{equation}
 r_0 = \frac{\hbar}{\sqrt {2 m_e \epsilon}}
\end{equation}
Putting these together we get
\begin{equation}
 \rho(l,l+1) = \frac{\hbar}{\sqrt {2 m_e \epsilon}} \frac{l+1}{2} \sqrt{\frac{2l+4}{2l+1}}
\end{equation}
Note that the applicability of our model for the bound electrons of water molecules requires that both the ground and excited states should be stable. This is not the case for free water molecule \cite{adhikari}. However the presence of the CNT boundary and the presence of the nearly free electrons within the CNT can  make the relevant electronic states stable so that coherent structures can form and the results used hold.

Using these ingredients, we find that the quantity $G$ has the form
\begin{equation}
 G = 9.3 \times 10^{-22} ~ \frac{l+1}{2} ~ \sqrt{\frac{2l+4}{2l+1}} ~ \frac{1}{r_e ~ \sqrt{E} ~ \sqrt{\epsilon}},
\end{equation}
where $r_e$ is measured in cm and the energies $E = \hbar \omega$ and $\epsilon$ are measured in ergs. Recall that the mixing of the two electronic states of water due to the ZPEM field increases the energy of the excited state by $G^2 E$ where $E = \hbar \omega$. In order for the system to be stable under dissociation, the condition $G^2 E < \epsilon$ has to be satisfied. This ensures that the inrease in the excited state energy due to the mixing caused by the ZPEM field does not lead to the ionization of the bound electron. This condition leads to the constraint $\epsilon > \epsilon_{min}$ where
\begin{equation}
 \epsilon_{min} = \frac{9.3 \times 10^{-22}}{r_e}~ \frac{l+1}{2} ~\sqrt{\frac{2l+4}{2l+1}}.
\end{equation}
For transition between $l=2$ and $l=3$ electonic states, and for the lowest effective radius $r_e = 2.33$ corresponding to the 1.05 nm CNT, we get the condition that $\epsilon_{min} = 0.05$ eV.

We are now ready to calculate the temperature of the coherent water layer. Substituting $\rho_{01}$ and $N$ in $G$ we get
\begin{equation}
 G = 9.3 \times 10^{-22} ~ \frac{l+1}{2} ~ \sqrt{\frac{2l+4}{2l+1}} ~ \frac{1}{r_e ~ \sqrt{E} ~ \sqrt{\epsilon}}
\end{equation}
Recall that for thermal stability, we need $kT < G^2 \hbar \omega$. Thus the maximum limiting temperature upto which the coherent structure could be thermally stable is given by $kT_{max} = G^2 E$, where $E = \hbar \omega$. From these conditions we get
\begin{equation}
 T_{max} = 62.67 \times 10^{-28} ~ \left ( \frac{l+1}{2} \right )^2 ~ \frac{2l+4}{2l+1} ~ \frac{1}{r_e^2 ~ \epsilon}.
\end{equation}
An immediate consequence of Eqn. (17) is that for two CNT's of effective radii $r_e^{[1]}$ and $r_e^{[2]}$, the corresponding transition temperatures are related by
\begin{equation}
 \frac{T_{max}(2)}{T_{max}(1)} = \frac{r_e^2 (1)}{r_e^2(2)}
\end{equation}
Thus the model predicts that the phase transition temperature ratio of water in two different CNT's is completely determined by the ratio of their effective radii, which is available for the water flow. It also predicts that the phase transition temperature decreases as the effective radius of the CNT increases. A comparison of the  prediction of this temperature ratio with that obtained in recent experiment \cite{mit} is given in Table 1. The data \cite{mit} further indicates that the ice like phase persists upto 30 C, which is approximately the room temperature. In our framework this corresponds to a CNT diameter of approximately 11.5 nm.

\begin{table}
\centering
    \begin{tabular} { | c | c | c | c | c | c| } 
\hline
$T_{max}^{(1.05)}$ (K) & $T_{max}^{(1.06)}$ (K)   & $\frac{T_{max}^{(1.05)}}{T_{max}^{(1.06)}}$  & $r_{e}^{(1.05)}$ ($A^{\circ}$) 
& $r_{e}^{(1.06)}$ ($A^{\circ}$) & $\left ( \frac{r_{e}^{(1.06)}}{r_{e}^{1.05}} \right )^2$\\ \hline
    424  & 390  & 1.09 & 2.33 & 2.38 & 1.04 \\ 
    \hline
\end{tabular}
\label{table:2}
\caption{The first three columns refer to the experimental data \cite{mit} for the 1.05 nm and 1.06 nm CNT's. We have used the maximum reported temperatures. The last three column show the result of our analysis. The ratios appearing in the third and sixth column give a direct comparison between the data and our theoretical prediction.}  
\end{table}

The value of $T_{max}$ depends on the angular momenta ($l,~ l+1)$ of the electronic states between which the transition takes place, $\epsilon$ and the effective radius $r_e$ of the coherent water structure in the CNT. We choose $l=2$ and $\epsilon = 0.06$ eV which satisfies the theoretical constraint that $\epsilon > 0.05$ eV. For the effective radius $r_e = 2.33$ A$^\circ$, corresponding to the 1.05 nm CNT, the model predicts the maximum phase transition temperature $T_{max} = 433$ K or equivalently 160 C. The maximum temperature signifies the upper limit beyond which the thermal instability sets in. Thus the actual temperature is expected to be somewhat less, which is consistent with the experimental data \cite{mit}. Once this choice is made, values of the maximum allowed solid-liquid phase transition temperatures for other effective radii can be obtained from Eqn. (18). Also note that as the CNT radius and equivalently the effective radius $r_e$ increases, the value of $G$ decreases. In our formalism this indicates that the mixing between the two electronic states of water becomes weaker and the coherent domains begin to disappear as the effective radius increases. The model thus predicts that stable coherent water cannot exist when the phase transition temperature is lower than the room temperature. 

\section{Conclusion}

The simple model proposed can satisfactorily explain the observed sensitive dependence of the phase-transition temperature on the CNT radius. It predicts that for $r_e > 11.5$ nm there is no ice like phase since the corresponding transition temperature is 303 K, approximately the room temperature. The model uses a simple geometric picture for air molecules to explain the effective radius size. Since the expulsion of air molecules from coherent water leading to the formation of a ring structure reduces the available volume for the water molecules. The model also predicts that the coherent phases of water appear as domains of length of 100-300 nm separated by a few molecules of air in a state of dynamic equilibrium. The key result is that the ratio of the transition temperatures. The presence of these air molecules should be amenable to detection by spectroscopic means.



\appendix 
\section{General expression for $G$}

The transition matrix element between the electronic bound states of water mixed by the zero point electrmagnetic (ZPEM) field is used to define the parameter $G$ as \cite{kumar}
\begin{eqnarray*}
 G \hbar \omega = <0|e \bf{x}.{\bf{\delta E}} |1>.
\end{eqnarray*}
Using $\frac{\bf{E}^2}{8 \pi} = \frac{\hbar \omega}{V}$ and $\delta {\bf{E}} = -{\bf{E}} \frac{\delta V}{2V}$, gives
\begin{eqnarray*}
 G=\sqrt{8\pi\alpha}~\sqrt{\frac{c}{V \omega}}~\frac{\delta V}{2V} \rho_{01},
\end{eqnarray*}
where $\rho_{01} \equiv <0|{\bf{x.u}}|1>$ \cite{kumar}. In vacuum, the fluctutaion of the E-M field with a frequency $\omega$ corresponds to a wavelength $\lambda$ such that $\omega \lambda = 2 \pi c$, where $c$ is the velocity of light in vacuum. In water, the same frequency is associated with a wavelength $\lambda^{\prime}$ such that
 $\lambda = \nu \lambda^{\prime}$ where $\nu$ is the refractive index of water.

The ZPEM field acts on all bound electrons in the mesoscopic volume $V$  to form a coherent domain of water. The mesoscopic volume is given by $V= \pi r_e^2 \lambda^{\prime}$. This gives $\frac{\delta V}{V} = 2 \frac{\delta r_e}{r_e}$, where 
\begin{eqnarray*}
 (\delta r_e)^2 = 2 N \frac{\alpha}{\pi} \lambda_c^2 \ln{\frac{1}{\alpha^2}}
\end{eqnarray*}
and $\lambda_c$ denotes the Compton wavelength \cite{welton} . From these we get
\begin{eqnarray*}
G &=& \sqrt{8\pi\alpha}~\sqrt{\frac{c}{V \omega}}~(\frac{\lambda_c}{r})~\sqrt{\frac{2\alpha }{\pi}~\ln{\frac{1}{\alpha^2}}}~\rho_{01}~\sqrt{N} \\
 &=& \frac{2 \sqrt{2}}{\pi}~\sqrt{\nu}~ \lambda_c ~\alpha ~ {\sqrt{\ln{\frac{1}{\alpha^2}}}} ~  ~ \frac{\rho_{01}~\sqrt{N}}{r^2}
\end{eqnarray*}
Usung $\alpha = \frac{1}{137}$ and $\lambda_c = 2.42 \times 10^{-10}$, we get
\begin{eqnarray*}
 G = 5 \times 10^{-12} \sqrt{\nu}~ \frac{\rho_{01} \sqrt{N}}{r_e^2} 
\end{eqnarray*}
The important feature of this formula is the dependence on the total number of coherent electrons $N$. We use simple geometric arguments to estimate the maximum allowed value of $N$.

\section{Expulsion of air molecules from water}

Dissolved air molecules in water are expelled when water is in a coherent phase. This happens because air molecules not vibrating in resonance with water molecules experience repulsive force as shown in \cite{frohlich}. The expelled air molecules form a ring around the water molecules preventing them in coming into direct contact with the carbon surface. Let us briefly give Frohlich's argument. Consider the following simple model in which a collection of free dipoles interact with an oscillating electromagnetic field. The interaction Hamiltonian $H$ is 
\begin{displaymath}
 H=g\int \psi^{*}(x,t)\psi(x,t) A(x,t)
\end{displaymath}
where $\psi^{*}(x,t)$ represents the dipole creation operator and $\psi(x,t)$ the dipole destruction operator and $A(x,t)$ the oscillating electromagnetic field
which we have approximated all fields by scalar fields. The oscillatory character of $A(x,t)$ means we can write it as
\begin{displaymath}
 A(x,t)=\frac{1}{\sqrt{V}} \sum [a_q e^{-1\omega_q t+\mathbf{q}.\mathbf{x}}+ hc]
\end{displaymath}
This gives rise to a positive interaction of the form
\begin{displaymath}
 V_q \approx \frac{\omega_q}{(\epsilon_p-\epsilon_{p+q})^2-(\omega_q)^2}
\end{displaymath}
where $\epsilon=\epsilon_p-\epsilon_{p+q}$ is  energy difference between the two dipoles and $\omega$ is the photon energy.
Integrating $\omega$ over the allowed momenta values of the photon
with an appropriate physically determined  cut off, $|Omega$,  gives the potential energy as a function of the difference in energies of the two dipoles. In the simple model we
take the dipoles to be free air molecules in the self generated gas cavity that surrounds a hydrophobic site. The quantum fluctuating momentum scale set by the quantum uncertainty principle is for a cavity size  of $ \Delta x \approx 1$ given by  $\Delta p \approx \frac{h}{\Delta x}$ . The corresponding energy is $\approx 10^{-4}$ eV.  Carrying out the momentum integration gives the dipole- photon potential $V_{d,\gamma}(\epsilon)$ to be
\begin{displaymath}
V_{d,\gamma}(\epsilon) \approx (\frac{g^2}{\Omega})(\frac{\epsilon}{2\Omega}\ln|\frac{1+\frac{\epsilon}{\Omega}}{1-\frac{\epsilon}{\Omega}}|-1)
\end{displaymath}
where $\epsilon=\epsilon_1-\epsilon_2$ the  difference of energies of the two dipoles and $\Omega$ is the photon energy corresponding to the thermal
 momentum cutoff momentum. We see this potential is attractive for
$\epsilon \leq \Omega$.  We note that this attractive potential, and its associated attractive force, increases as $\epsilon \Rightarrow 0$ which corresponds to the oscillators being in resonance. This new force, of quantum origin, only appears when the dipoles are oscillating and the electromagnetic field is time dependent. The oscillation of the dipoles in the model come from
the fact that they are confined in a bubble cavity of small length and 
are represented by oscillating waves.  This new quantum force  brings resonating dipoles close together and repels ones that are not in resonance as stated.

\end{document}